\documentclass{ifacconf}

\usepackage{graphicx}      
\usepackage{natbib}        

\usepackage{url}
\usepackage{amsmath}
\usepackage{amsfonts}
\usepackage{graphicx}
\usepackage{amssymb}
\usepackage{alltt}
\usepackage{listings}
\usepackage[USenglish]{babel}
\usepackage{caption}
\usepackage{subcaption}
\begin{document}
\begin{frontmatter}

\title{Linear Modeling of a Flexible Substructure Actuated through Piezoelectric Components for Use in Integrated Control/Structure Design} 


\author[First]{J. Alvaro PEREZ} 
\author[Second]{Thomas LOQUEN} 
\author[Third]{Daniel ALAZARD} 
\author[Fourth]{Christelle PITTET}

\address[First]{ONERA System Control and Flight Mechanics Department, 
   Toulouse, 31000 France (e-mail: Jose-Alvaro.Perez\_ Gonzalez@onera.fr}
   \address[Second]{ONERA System Control and Flight Mechanics Department, 
   Toulouse, 31000 France (e-mail: thomas.loquen@onera.fr}
   \address[Third]{ISAE System Control Department, 
   Toulouse, 31000 France (e-mail: daniel.alazard@isae.fr}
      \address[Fourth]{CNES AOCS Department, 
   Toulouse, 31000 France (e-mail: christelle.pittet@cnes.fr}

\begin{abstract}                
\quad This study presents a generic TITOP (Two-Input Two-Output Port) model of a substructure actuated with embedded piezoelectric materials as actuators (PEAs), previously modeled with the FE technique. This allows intuitive assembly of actuated flexible substructures in large flexible multi-body systems. The modeling technique is applied to an illustrative example of a flexible beam with bonded piezoelectric strip and vibration attenuation of a chain of flexible beams.
\end{abstract}


\end{frontmatter}

\section{Introduction}

Piezoelectric actuators and sensors (PEAs) have been widely used in the field of system control design of large flexible structures. However, the design of control systems involving PEAs requires an accurate knowledge of the electro-mechanical behaviour of the system for vibration dynamics, transfers between the inputs and the outputs and non-linear effects such as hysteresis and creep effect. In order to integrate PEAs in the controlled structure, a design procedure including virtual prototyping of piezoelements integrated with the structure needs to be developed. 

Macroscopic PEAs models  are divided in two main categories. In the first category, the behaviour of a PEAs is decoupled in several contributions such as hysteresis, vibration dynamics and creep based on physical laws. The most well-known model structure of PEAs is the electro-mechanical model proposed in \cite{Goldfarb1997_PEA}, in which all effects are taken into account. In this category, other models only consider vibration dynamics with Finite Element (FE) models (\cite{Piefort2000_PEA}) or static behaviour (\cite{Smits1991_PEA}). The second category does not decouple the different behaviours of the PEA, all effects are considered simoultaneously. However, they are only accurate over small frequency ranges, what seriously limit their usage.

This study presents a PEAs modelisation technique that allows considering piezoelectric actuated flexible substructures linked with other substructures. Based on the TITOP approach explained in \cite{Perez2015_LM}, \cite{Alazard2015_LM} and \cite{Perez_IFAtheory2015}, the method casts in state-space form the FE model of an actuated flexible substructure in order to consider the acceleration-loads transfer within a flexible-multibody system. Section \ref{sec:finite element} introduces the main equations of a piezoelectric FE model. then, Section \ref{sec:TITOPmodeling} explains how to obtain the Two-Input Two-Output Port(TITOP) model through Component Modes Synthesis (CMS) and the double-port approach. Then, an application for a beam with bonded piezoelectric strip is illustrated. Finally, application to vibration attenuation of a chain of flexible beams is performed and conclusions are stated.

\section{Finite Element Modeling of a Piezoelectric Component}
\label{sec:finite element}

As stated in \cite{Ansi1988_PEA}, the constitutive linear equations of an element piezoelectric material read:

\begin{equation}
\begin{Bmatrix} T \end{Bmatrix} = \begin{bmatrix} c^E
\end{bmatrix}\begin{Bmatrix} S \end{Bmatrix} - \begin{bmatrix} e \end{bmatrix}^T \begin{Bmatrix} E \end{Bmatrix}
\end{equation} 

\begin{equation}
\begin{Bmatrix} D \end{Bmatrix} = \begin{bmatrix} e
\end{bmatrix}\begin{Bmatrix} S \end{Bmatrix} + \begin{bmatrix} \epsilon^S \end{bmatrix}\begin{Bmatrix} E \end{Bmatrix}
\end{equation} 

where $\begin{Bmatrix}T\end{Bmatrix} $ is the stress vector, $\begin{Bmatrix} S \end{Bmatrix}$ the deformation vector,$ \begin{Bmatrix} E \end{Bmatrix}$ the electric field, $ \begin{Bmatrix} D \end{Bmatrix}$ the electric displacement, $ \begin{bmatrix} c \end{bmatrix}$ the elasticity constants matrix, $ \begin{bmatrix} \epsilon \end{bmatrix}$ the dielectric constants matrix, $ \begin{bmatrix} e \end{bmatrix}$ the piezoelectric constants, with superscripts $^E$, $^S$ and $^T$ indicating static conditions for $E$, $S$ and $T$ respectively.

The dynamic equations of a piezoelectric continuum can be discretized in elements and written in the finite element formulation as follows:

\begin{equation}
\begin{bmatrix} \mathcal{M}_{qq} \end{bmatrix} \begin{Bmatrix} \ddot{q} \end{Bmatrix} + \begin{bmatrix} \mathcal{K}_{qq} \end{bmatrix} \begin{Bmatrix} q \end{Bmatrix} + \begin{bmatrix} \mathcal{K}_{q\phi} \end{bmatrix} \begin{Bmatrix} \phi \end{Bmatrix} = \begin{Bmatrix} f \end{Bmatrix}
\end{equation}

\begin{equation}
\begin{bmatrix} \mathcal{K}_{\phi q} \end{bmatrix} \begin{Bmatrix} q \end{Bmatrix} + \begin{bmatrix} \mathcal{K}_{\phi \phi} \end{bmatrix} \begin{Bmatrix} \phi \end{Bmatrix} = \begin{Bmatrix} \gamma \end{Bmatrix}
\end{equation}

where the element coordinates $\begin{Bmatrix} q \end{Bmatrix}$, the applied voltage $\begin{Bmatrix} \phi \end{Bmatrix}$, the electric charge $\begin{Bmatrix} \gamma \end{Bmatrix}$ and external forces $\begin{Bmatrix} f \end{Bmatrix}$  are related through the element mass matrix, $\begin{bmatrix} \mathcal{M}_{qq} \end{bmatrix}$, the element stiffness matrix, $\begin{bmatrix} \mathcal{K}_{qq} \end{bmatrix}$, the piezoelectric coupling matrix $\begin{bmatrix} \mathcal{K}_{q\phi} \end{bmatrix}$ and the capacitance matrix $\begin{bmatrix} \mathcal{K}_{\phi \phi} \end{bmatrix}$. Upon carrying out the assembly of each piezoelectric element, we get  the global system of equations:

\begin{equation}
\label{eq:generalPiezo1}
\begin{bmatrix} M_{uu} \end{bmatrix} \begin{Bmatrix} \ddot{u} \end{Bmatrix} + \begin{bmatrix} K_{uu} \end{bmatrix} \begin{Bmatrix} u\end{Bmatrix} + \begin{bmatrix} K_{uv} \end{bmatrix} \begin{Bmatrix} v \end{Bmatrix} = \begin{Bmatrix} F \end{Bmatrix}
\end{equation}

\begin{equation}
\label{eq:generalPiezo2}
\begin{bmatrix} K_{v u} \end{bmatrix} \begin{Bmatrix} u \end{Bmatrix} + \begin{bmatrix} K_{vv} \end{bmatrix} \begin{Bmatrix} v \end{Bmatrix} = \begin{Bmatrix} g \end{Bmatrix}
\end{equation}

where the global coordinates $\begin{Bmatrix} u \end{Bmatrix}$, the global applied voltage $\begin{Bmatrix} v \end{Bmatrix}$, the electric charge $\begin{Bmatrix} g \end{Bmatrix}$ and external forces $\begin{Bmatrix} F \end{Bmatrix}$ are now related through the global  mass matrix, $\begin{bmatrix} M_{uu} \end{bmatrix}$, the global stiffness matrix, $\begin{bmatrix} K_{uu} \end{bmatrix}$, the piezoelectric coupling matrix $\begin{bmatrix} K_{uv} \end{bmatrix}$ and the capacitance matrix $\begin{bmatrix} K_{vv} \end{bmatrix}$.

\section{TITOP Modeling of the Piezoelectric Component}
\label{sec:TITOPmodeling}

The Two-Input Two-Output Port (TITOP) model of the piezoelectric component is obtained through the application of the Component Modes Synthesis (CMS) transformation to Eqns. (\ref{eq:generalPiezo1}) and (\ref{eq:generalPiezo2}), then casting the resulting transformation into a state-space representation for the desired inputs-outputs.

\subsection{Component Modes Synthesis Transformation}
\label{subsec:CMS}

The Component Modes Synthesis transformation allows separating the different contributions of elastic body displacements into rigid body, redundant boundaries and internal elastic displacements. The resulting equations are easier to manipulate since rigid-body and elastic displacements appear uncoupled in the transformed stiffness matrix. The fundamentals of Component Modes Synthesis were stated by \cite{Hurty1965_CM} in 1965 and then recalled later by \cite{Craig2000_CB}, the reader might head to those references if more information about CMS is desired.

The global coordinates $\begin{Bmatrix} u \end{Bmatrix}$ are then partitioned into three main sets: rigid-body coordinates, $r$, redundant boundary coordinates, $c$, and fixed-constraint coordinates, $n$. Applying this division, Eqns. (\ref{eq:generalPiezo1}) and (\ref{eq:generalPiezo2}) result:

\small
\begin{equation}
\begin{split}
\label{eq:generalPiezo1divided}
& \begin{bmatrix} M_{nn} & M_{nc} & M_{nr} \\ M_{cn} & M_{cc} & M_{cr} \\ M_{rn} & M_{rc} & M_{rr}\end{bmatrix}_{uu} \begin{Bmatrix} \ddot{u}_n \\ \ddot{u}_c \\ \ddot{u}_r  \end{Bmatrix} + \begin{bmatrix} K_{nn} & K_{nc} & K_{nr} \\ K_{cn} & K_{cc} & K_{cr} \\ K_{rn} & K_{rc} & K_{rr}\end{bmatrix}_{uu} \begin{Bmatrix} u_n \\ u_c \\ u_r  \end{Bmatrix} + \\
& +\begin{bmatrix} K_{nv} \\ K_{cv} \\ K_{rv} \end{bmatrix} \begin{Bmatrix} v \end{Bmatrix} = \begin{Bmatrix} F_n \\ F_c + \tilde{F}_c \\ F_r + \tilde{F}_r \end{Bmatrix}
\end{split}
\end{equation}

\begin{equation}
\label{eq:generalPiezo2divided}
\begin{bmatrix} K_{v n} & K_{v c} & K_{v r}  \end{bmatrix} \begin{Bmatrix} u_n \\ u_c \\ u_r  \end{Bmatrix} + \begin{bmatrix} K_{vv} \end{bmatrix} \begin{Bmatrix} v \end{Bmatrix} = \begin{Bmatrix} Q \end{Bmatrix}
\end{equation}
\normalsize

where $\begin{bmatrix} M \end{bmatrix}$, $ \begin{bmatrix} K \end{bmatrix}$, $\begin{Bmatrix} u \end{Bmatrix}$ and  $\begin{Bmatrix} F \end{Bmatrix}$ have been partitioned into their contributions to rigid-body, redundant boundaries and fixed-boundary displacements. The ``tilde'' load term, $\tilde{F}_r$ and $\tilde{F}_c$, denotes the force resulting from the connection to adjacent structures at the boundary points.

In CMS, physical displacements can be expressed in terms of generalized coordinates by the Rayleigh-Ritz coordinate transformation \cite{Craig2000_CB}:

\begin{equation}
\label{eq:phi}
\begin{Bmatrix} u_n\\ u_c \\u_r \end{Bmatrix} = \begin{bmatrix}\Phi \end{bmatrix} \begin{Bmatrix} \eta_n\\ \eta_c \\ \eta_r \end{Bmatrix}  = \begin{bmatrix}
\phi_{nn} & \phi_{nc} & \phi_{nr} \\ 0_{cn} & I_{cc} & \phi_{cr} \\ 0_{rn} & 0_{rc} & I_{rr}
\end{bmatrix} \begin{Bmatrix} \eta_n\\ \eta_c \\ \eta_r \end{Bmatrix}
\end{equation}

where the component-mode matrix $\begin{bmatrix}
\Phi \end{bmatrix}$ is a matrix of preselected
component modes including: fixed-constraint modes,$n$,
redundant boundary modes, $c$, and rigid-body modes, $r$. Pre-multiplying by $\begin{bmatrix}
\Phi \end{bmatrix}^T$, substituting Eqn. (\ref{eq:phi}) into Eqns.  (\ref{eq:generalPiezo1divided}) and (\ref{eq:generalPiezo2divided}) and considering that neither interior forces nor external forces apply ($F_n=F_c=F_r=0$), Eqn. (\ref{eq:generalPiezo1divided}) yields:

\small
\begin{equation}
\begin{split}
\label{eq:generalPiezo1normed}
& \begin{bmatrix} \hat{M}_{nn} & \hat{M}_{nc} & \hat{M}_{nr} \\ \hat{M}_{cn} & \hat{M}_{cc} & \hat{M}_{cr} \\ \hat{M}_{rn} & \hat{M}_{rc} & \hat{M}_{rr}\end{bmatrix}_{\eta\eta} \begin{Bmatrix} \ddot{\eta}_n \\ \ddot{\eta}_c \\ \ddot{\eta}_r  \end{Bmatrix} + \begin{bmatrix} \hat{K}_{nn} & 0_{nc} & 0_{nr} \\ 0_{cn} & \hat{K}_{cc} & 0_{cr} \\ 0_{rn} & 0_{rc} & 0_{rr}\end{bmatrix}_{\eta\eta} \begin{Bmatrix} \eta_n \\ \eta_c \\ \eta_r  \end{Bmatrix} + \\ & + \begin{bmatrix}\hat{K}_{nv} \\ \hat{K}_{cv}  \\ \hat{K}_{rv}  \end{bmatrix} \begin{Bmatrix} v \end{Bmatrix} = \begin{Bmatrix} 0\\ \tilde{F}_c \\ \tilde{F}_r + \phi_{cr}^T \tilde{F}_c \end{Bmatrix}
\end{split}
\end{equation}

\begin{equation}
\label{eq:generalPiezo2normed}
\begin{bmatrix} \hat{K}_{nv} &\hat{K}_{cv}  & \hat{K}_{rv}  \end{bmatrix} \begin{Bmatrix} \eta_n \\ \eta_c \\ \eta_r  \end{Bmatrix} + \begin{bmatrix} K_{vv} \end{bmatrix} \begin{Bmatrix} v \end{Bmatrix} = \begin{Bmatrix} G \end{Bmatrix}
\end{equation}
\normalsize

with the new coupling matrix coefficients:
\begin{equation}
\hat{K}_{nv} = \phi_{nn}K_{nv}
\end{equation}
\begin{equation}
\hat{K}_{cv} = K_{cv} + \phi_{cn}K_{nv} 
\end{equation}
\begin{equation}
\hat{K}_{rv} = K_{rv} + \phi_{rc}K_{cv} + \phi_{rn}K_{nv}
\end{equation}

Equations (\ref{eq:generalPiezo1normed}) and (\ref{eq:generalPiezo2normed}) are the most generalized expression for a FE model of a piezoelectric component transformed through the CMS method. Section \ref{subsec:SS} will show how to take advantage of this form in order to simply model an accurate piezoelectric component for control of flexible multi-body systems.

\subsection{Actuated TITOP State-Space Realization}
\label{subsec:SS}

The purpose of a TITOP state-space realization of Eqns. (\ref{eq:generalPiezo1normed}) and (\ref{eq:generalPiezo2normed}) is to model the piezoelectric component as a substructure connected with two different structures in chain-like assembly through two connection points, $P$ and $Q$. This is useful in order to consider the actuation provoked by the piezoelectric component to the rest of the structure, modeled as a flexible multibody system.

\begin{figure} 
\centering
\includegraphics[scale=0.75]{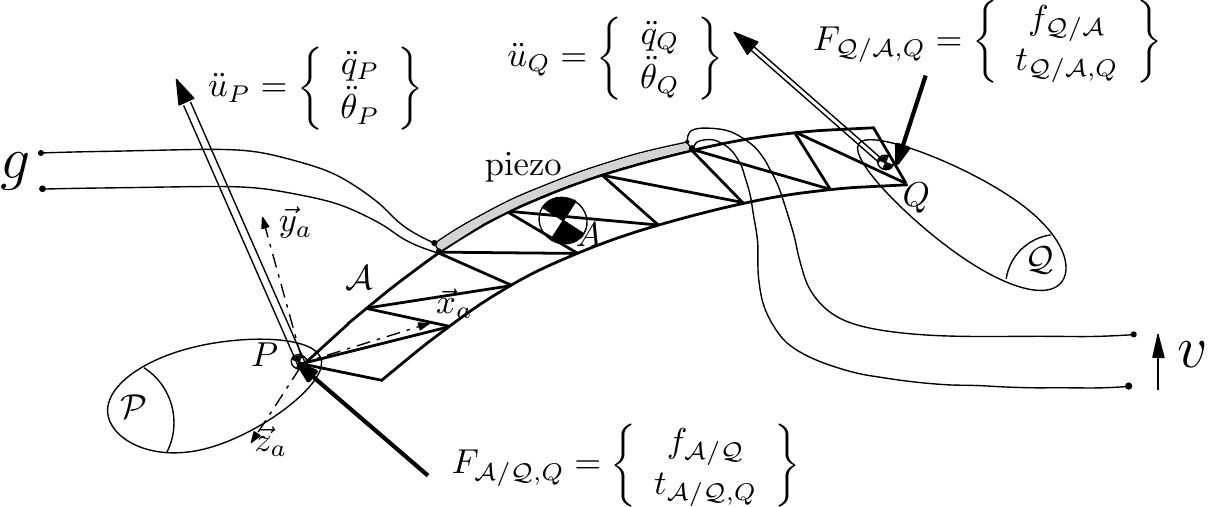}
\caption{Illustration of a generic substructure with bonded piezoelectric material}
\label{fig:genericPiezo}
\end{figure}

As shown in Fig. \ref{fig:genericPiezo}, the flexible piezoelectric component $\mathcal{A}$ is linked to the parent structure $\mathcal{P}$ at the point $P$ and to a child substructure $\mathcal{Q}$ at the point $Q$. It is assumed that the only external loads applied to $\mathcal{A}$ are the interactions  with $\mathcal{P}$  at point $P$ and with $\mathcal{Q}$ at point $Q$, as hypothesis on Eqn. (\ref{eq:generalPiezo1normed}) states. Voltage $v$ can be applied to the piezo in order to provoke a electric field, and the electric charge $g$ is an available measure.

The main problem is how to consider the electro-mechanical coupling between  $\mathcal{P}$, $\mathcal{A}$, $\mathcal{Q}$ and electric states.  The mechanical overlapping between substructures is expressed as an acceleration-load transfer through the common boundaries, called the \textbf{double-port approach}, proposed by \cite{Alazard2015_LM} and \cite{perez_IFAtheory2015}. With this approach, both points, $P$ and $Q$, suffer an acceleration-load transfer, in such a way that the acceleration is transferred to the next substructure in the chain ($\mathcal{Q}$ in this case) and the load is transmitted to the previous substructure  in the chain ( the parent $\mathcal{P}$ structure). 

A generalization of the double-port approach is presented in this study for the piezoelectric FE CMS transformation. The existing electro-mechanical coupling between loads-accelerations and voltage-charge is considered through the augmentation of the classic double-port model with an additional electric input, the applied voltage $v$, and with an additional output, the electric charge, $g$, such that:

\begin{equation}\label{eq:MAPC}
\begin{Bmatrix}\ddot{u}_Q \\ F_{\mathcal{A/P},P} \\ g \end{Bmatrix}= \begin{bmatrix}G^{\mathcal{A}}_{P,Q}(s) \end{bmatrix} \begin{Bmatrix} F_{\mathcal{Q/A},Q} \\\ddot{u}_P \\ v \end{Bmatrix}
\end{equation}

Therefore Eqn. (\ref{eq:MAPC}) relates the accelerations suffered at connection point $P$, loads at connection point $Q$ and the applied voltage $v$ to the acceleration at connection point $Q$, transmitted force to the previous substructure $\mathcal{P}$ and the measured electric charge $g$. An assignment of the different degrees of freedom is performed in order to distribute the acceleration-load transfer: rigid-body displacements are those of connection point $P$ and the redundant constraint displacements those of connection point $Q$. Thus the accelerations read:

\begin{equation}
\label{eq:PandQacc}
\begin{split}
& \ddot{u}_P = \ddot{\eta}_r \\ 
&\ddot{u}_Q = \ddot{\eta}_c + \phi_{cr}\ddot{\eta}_r
\end{split}
\end{equation}

Equation (\ref{eq:PandQacc}) implies that the rigid motion is supported by point $P$ and the constrained motion of connection point $Q$ is a result of the rigid body motion in $P$ transported to point $Q$ ($\phi_{cr}\ddot{\eta}_r$) plus the constrained motion due to flexibility ($\ddot{\eta}_c$).

In the same way, loads are received and transmitted by appendage $\mathcal{A}$ with the following directions:

\begin{equation}
\label{eq:PandQfor}
\begin{split}
&F_{\mathcal{A}/\mathcal{P},P}  = - \tilde{F}_r   \\
& F_{\mathcal{Q}/\mathcal{A},Q} =   \tilde{F}_c  
\end{split}
\end{equation}

Using the relations given in Eqns. (\ref{eq:PandQacc}) and (\ref{eq:PandQfor}) in combination with Eqn. (\ref{eq:generalPiezo1normed}), a state-space representation can be obtained for the piezoelectric component $\mathcal{A}$:

\begin{eqnarray}
\label{eq:TITOP}
G^{\mathcal{A}}_{P,Q}(s)
\left\lbrace
\begin{array}{l}
\begin{Bmatrix}
\dot{\eta}_n\\ 
\dot{\eta}_c \\ 
\ddot{\eta}_n\\ 
\ddot{\eta}_c 
\end{Bmatrix}  = A
\begin{Bmatrix}
\eta_n\\ 
\eta_c \\
\dot{\eta}_n\\ 
\dot{\eta}_c 
\end{Bmatrix} + B \begin{Bmatrix} F_{\mathcal{Q}/\mathcal{A},Q} \\ 
\ddot{u}_P \\ v \end{Bmatrix} \\ \\
\begin{Bmatrix} \ddot{u}_Q  \\  F_{\mathcal{A}/\mathcal{P},P} \\ g_c \end{Bmatrix}= C \begin{Bmatrix}
\eta_n\\ 
\eta_c \\
\dot{\eta}_n\\ 
\dot{\eta}_c 
\end{Bmatrix} + (D + D_\delta )\begin{Bmatrix} F_{\mathcal{Q}/\mathcal{A},Q} \\ 
\ddot{u}_P \\ v \end{Bmatrix} \end{array} \right.
\end{eqnarray}

where $A$, $B$, $C$, $D$ and $D_\delta$ are the short hand notation of the following state-space matrices:

\begin{equation} \label{eq:ATITOP}
A =\left[ \begin{array}{cc}
0_{n+c} & I_{n+c} \\ 
- \hat{M}_Q^{-1} \hat{K}_Q & - \hat{M}_Q^{-1} \hat{D}_Q  
\end{array} \right]
\end{equation}

\begin{eqnarray} \label{eq:BTITOP}
B = \begin{bmatrix}
0_{n+c,c+r+v} \\
\hat{M}_Q^{-1} \begin{bmatrix}
0_{nc}& -\hat{M}_{nr} & -\hat{K}_{nv} \\
I_{cc}& -\hat{M}_{cr} & -\hat{K}_{cv}
\end{bmatrix}
\end{bmatrix} 
\end{eqnarray}

\begin{eqnarray} \label{eq:CTITOP}
C = \begin{bmatrix}
\begin{bmatrix}
0_{cn} & I_{cc} \end{bmatrix} \begin{bmatrix} -\hat{M}_Q^{-1} \hat{K}_Q
& -\hat{M}_Q^{-1} \hat{D}_Q
\end{bmatrix} \\
\begin{bmatrix}
\hat{M}_{rn} & \hat{M}_{rc}\end{bmatrix} \begin{bmatrix}\hat{M}_Q^{-1} \hat{K}_Q
& \hat{M}_Q^{-1} \hat{D}_Q
\end{bmatrix}\\
\begin{bmatrix}
\hat{K}_{vn} & \hat{K}_{vc}\end{bmatrix} \begin{bmatrix} I_{n+c,n+c} & 0_{n+c,n+c}
\end{bmatrix}
\end{bmatrix} 
\end{eqnarray}

\begin{eqnarray} \label{eq:DTITOP}
D = \begin{bmatrix}
\begin{bmatrix}
0_{cn} & I_{cc} \end{bmatrix}  \hat{M}_Q^{-1} \begin{bmatrix}
0_{nc}& -\hat{M}_{nr} & -\hat{K}_{nv} \\
I_{cc}& -\hat{M}_{cr} & -\hat{K}_{cv}
\end{bmatrix} \\
\begin{bmatrix}
\hat{M}_{rn} & \hat{M}_{rc}\end{bmatrix} \hat{M}_Q^{-1} \begin{bmatrix}
0_{nc}& -\hat{M}_{nr} & -\hat{K}_{nv} \\
I_{cc}& -\hat{M}_{cr} & -\hat{K}_{cv}
\end{bmatrix} \\
0_{vc} \quad 0_{vr} \quad K_{vv}
\end{bmatrix} 
\end{eqnarray}

with 

\begin{equation} \label{eq:DdTITOP}
\begin{split}
& \hat{M}_Q =
\begin{bmatrix}
\hat{M}_{nn} & \hat{M}_{nc} \\
\hat{M}_{cn} & \hat{M}_{cc}
\end{bmatrix} ; \; \hat{K}_Q =
\begin{bmatrix}
\hat{K}_{nn} & 0_{nc} \\
0_{cn} & \hat{K}_{cc}
\end{bmatrix}; \; \\ 
&\hat{D}_Q =
\begin{bmatrix}
\hat{D}_{nn} & 0_{nc} \\
0_{cn} & \hat{D}_{cc}
\end{bmatrix}; \; D_{\delta} = \begin{bmatrix} 0_{cc} & \phi_{cr} & 0_{cv}\\ 
-\phi_{cr}^T & \hat{M}_{rr} & \hat{K}_{rv} \\ 0_{vc} & 0_{vr} & 0_{vv}
\end{bmatrix} ;
\end{split}
\end{equation}

\begin{figure} 
\center
\includegraphics[scale=0.5]{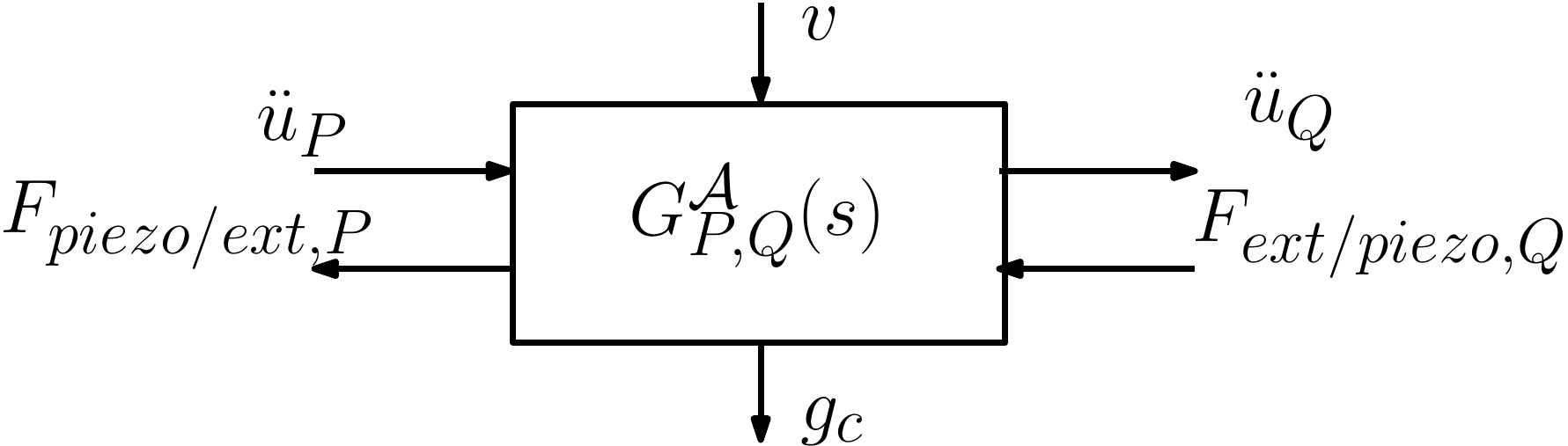}
\caption{Block diagram of TITOP modeling of a substructure actuated with piezoelectric materials}
\label{fig:pztTITOP}
\end{figure}

Equation (\ref{eq:TITOP}) with Eqns. (\ref{eq:ATITOP}), (\ref{eq:BTITOP}), (\ref{eq:CTITOP}), (\ref{eq:DTITOP}) and (\ref{eq:DdTITOP}) form the double-port model, $ \begin{bmatrix} G^\mathcal{A}_{P,Q} (s) \end{bmatrix} $, of the flexible piezoelectric component $\mathcal{A}$ in chain-like assembly, called \textbf{actuated Two-Input Two-Output Port (TITOP)} model. This model allows to interconnect different flexible substructures in chain-like assembly taking into account flexible motions. A simplified scheme of the TITOP model is shown in Fig. \ref{fig:pztTITOP}. In the 6 degrees of freedom case with one piezoelectric strip, $ \begin{bmatrix} G^\mathcal{A}_{P,Q} (s) \end{bmatrix} $, of the flexible substructure $\mathcal{A}$ is a $13 \times 13$ transfer matrix (that is, $r=6$, $v=1$, $c=6$, $g=1$). It should be noticed that the measured charge is denoted $g_c$ and not $g$ because only relative displacement between connection point $P$ and connection point $Q$ is measured for the sake of input/output simplicity.

The physical interpretation of Eqn. (\ref{eq:TITOP}) is as follows. The rigid-body displacements of the appendage $\mathcal{A}$ are transmitted by its connection point $P$ through the whole of the appendage, and this excites the fixed-boundary natural modes (the modes obtained when clamping the appendage at point $P$ and $Q$) through the modal participation matrices, $\begin{bmatrix} \hat{M}_{rn} \end{bmatrix}$ and $\begin{bmatrix}\hat{M}_{rc} \end{bmatrix}$, and thus the constraint point $Q$. These natural modes produce a load transmitted to substructure $\mathcal{P}$ modifying the load that appendage $\mathcal{A}$ will induce to $\mathcal{P}$, which depends on the load received at point $Q$, $F_{\mathcal{Q}/\mathcal{A},P}$, the acceleration received at point $P$, $\begin{Bmatrix} \ddot{u}_P \end{Bmatrix}$ and the natural modes. In addition, voltage $v$ modifies these transfers by increasing or reducing the transferred loads at $P$ and $Q$, and the measured electric charge $g$ responds to displacements changes either by mechanical interaction (through load application) or by electric field application (voltage).

\section{Modeling Application}
\label{sec:modeling}

\begin{figure} 
\center
\includegraphics[scale=0.65]{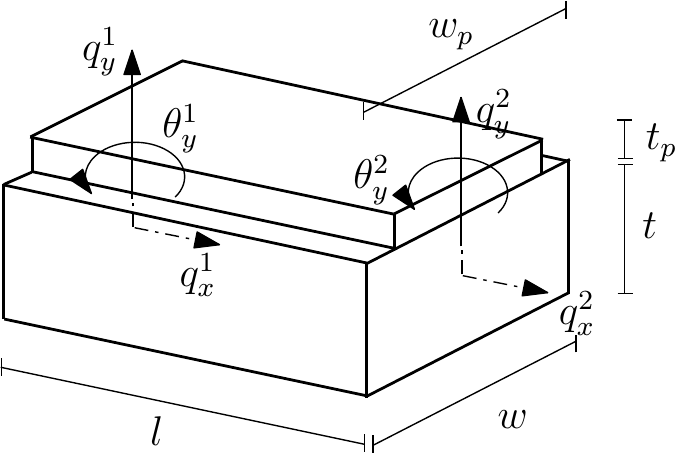}
\caption{Beam element with bonded piezoelectric strap }
\label{fig:beamPiezoElement}
\end{figure}

As an illustrative example, let consider a beam linking two substructures at both ends with a piezoelectric strip which has been appropriately bonded to one side, as seen in Fig. \ref{fig:beamPiezoElement}. The thickness of the piezoelectric strip is $t_p$, with a width denoted by $w_p$, Young's modulus $E_p$, density $\rho_p$ and cross-section inertia $I_p$. The piezoelectric strip is used
as an actuator by controlling the voltage $v$ applied to the electrodes‚ creating a constant electric field $E_3 = v/t_p$ across the thickness of the laminate.

The beam is modeled with a classic FE decomposition in several beam elements of length $l$. The beam has the same geometrical properties as the piezoelectric strip but denoted without the $_p$ subindex: $t$,$w$,$E$,$\rho$ and $I$. Since the piezoelectric laminate is glued to the beam, the mass and stiffness matrices of a beam element can be obtained as a sum of the contributions of the beam and the piezoelectric material. Thus:

\begin{equation}
\label{eq:mass}
\begin{bmatrix}
\mathcal{M}_{qq}
\end{bmatrix} = \begin{bmatrix}
\mathcal{M}_{qq}
\end{bmatrix}_{beam} +\begin{bmatrix}
\mathcal{M}_{qq}
\end{bmatrix}_{piezo} 
\end{equation}

\begin{equation}
\label{eq:stiffness}
\begin{bmatrix}
\mathcal{K}_{qq}
\end{bmatrix} = \begin{bmatrix}
\mathcal{K}_{qq}
\end{bmatrix}_{beam} +\begin{bmatrix}
\mathcal{K}_{qq}
\end{bmatrix}_{piezo}
\end{equation}

The matrices of Eqns. (\ref{eq:mass}) and (\ref{eq:stiffness}) are detailed in Appendix \ref{sec:appendix1}, obtained with classical FE decomposition with beam elements. With further FE decomposition to the piezoelectric strip, the following coupling matrix  and capacitance matrix are obtained:

\begin{equation}
\begin{bmatrix}
\mathcal{K}_{qv}
\end{bmatrix} = \begin{bmatrix}
-d_{31}E_pw_p \\ 0 \\ -d_{31}E_pw_p(t+\frac{t_p}{2} )\\ d_{31}E_pw_p \\ 0 \\ d_{31}E_pw_p(t+\frac{t_p}{2}) 
\end{bmatrix}
\end{equation}

\begin{equation}
\begin{bmatrix}
\mathcal{K}_{vv}
\end{bmatrix} = \frac{w_pl}{t_p}(\epsilon^T_{33} - d_{31}^2E_p)
\end{equation}

where $d_{31}$ is the piezoelectric constant under constant stress which relates the shrinkage observed in the direction $1$ when an electric field $E_3$ is applied along the direction of polarization $3$. The coefficient $\epsilon^T_{33}$ is the dielectric constant of the material. 

The global mass, stiffness, coupling and capacitance matrices are obtained by appending the elements matrices along the diagonal. Common node points between two points are simply added together. As an example the casting of stiffness matrices is given bellow for a $2\times2$ case:

\begin{equation}
\begin{split}
\begin{bmatrix}
& \mathcal{K}_{qq}
\end{bmatrix}^1 = \begin{bmatrix}
k^1_{11} & k^1_{12} \\ k^1_{21} & k^1_{22}
\end{bmatrix}; \quad \begin{bmatrix}
\mathcal{K}_{qq}
\end{bmatrix}^2 = \begin{bmatrix}
k^2_{11} & k^2_{12} \\ k^2_{21} & k^2_{22}
\end{bmatrix} \rightarrow  \\ \rightarrow \begin{bmatrix}
K_{uu}
\end{bmatrix} = \begin{bmatrix}
k^1_{11} & k^1_{12} & 0\\ k^1_{21} & k^1_{22}+ k^2_{11} & k^2_{12}  \\ 0 & k^2_{21} & k^2_{22} 
\end{bmatrix}
\end{split}
\end{equation}

Obtaining the global piezoelectric coupling matrix is more complicated since it depends on the desired profile of voltages for the piezoelectric strips. If the same voltage is applied among all the strips then the assembly is straightforward:

\begin{equation}
\begin{bmatrix}
\mathcal{K}_{qv}
\end{bmatrix}^1 = \begin{bmatrix}
k^1_{1v} \\  k^1_{2v}
\end{bmatrix}; \quad \begin{bmatrix}
\mathcal{K}_{qv}
\end{bmatrix}^2 = \begin{bmatrix}
k^2_{1v} \\  k^2_{2v}
\end{bmatrix} \rightarrow \begin{bmatrix}
K_{uv}
\end{bmatrix} = \begin{bmatrix}
k^1_{1v} \\  k^1_{2v}+ k^2_{1v}  \\ k^2_{2v} 
\end{bmatrix}
\end{equation}

\begin{equation}
\begin{bmatrix}
\mathcal{K}_{vv}
\end{bmatrix}^1 = \begin{bmatrix}
k^1_{vv} 
\end{bmatrix}; \quad \begin{bmatrix}
\mathcal{K}_{vv}
\end{bmatrix}^2 = \begin{bmatrix}
k^2_{vv}
\end{bmatrix} \rightarrow \begin{bmatrix}
K_{vv}
\end{bmatrix} = \begin{bmatrix}
  k^1_{vv}+ k^2_{vv} 
\end{bmatrix}
\end{equation}

However, if a different voltage is applied for each strip on each element then one has:

\begin{equation}
\begin{bmatrix}
\mathcal{K}_{qv_1}
\end{bmatrix}^1 = \begin{bmatrix}
k^1_{1v_1} \\  k^1_{2v_1}
\end{bmatrix}; \quad \begin{bmatrix}
\mathcal{K}_{qv_2}
\end{bmatrix}^2 = \begin{bmatrix}
k^2_{1v_2} \\  k^2_{2v_2}
\end{bmatrix} \rightarrow \begin{bmatrix}
K_{uv}
\end{bmatrix} = \begin{bmatrix}
k^1_{1v_1} & 0 \\   k^1_{2v_1} & k^2_{1v_2}  \\ 0 & k^2_{2v_2} 
\end{bmatrix}
\end{equation}

\begin{equation}
\begin{bmatrix}
\mathcal{K}_{vv}
\end{bmatrix}^1 = \begin{bmatrix}
k^1_{vv} 
\end{bmatrix}; \quad \begin{bmatrix}
\mathcal{K}_{vv}
\end{bmatrix}^2 = \begin{bmatrix}
k^2_{vv} 
\end{bmatrix} \rightarrow \begin{bmatrix}
K_{vv}
\end{bmatrix} = \begin{bmatrix}
k^1_{vv} &  0  \\ 0 & k^2_{vv} 
\end{bmatrix}
\end{equation}

In this last case, the applied voltage is a vector containing the different voltages applied along the assembled beam, whereas in the first case voltage is a scalar. The second option might be more suitable when more effectiveness of the actuator is required.

\begin{figure} 
\center
\includegraphics[scale=0.3]{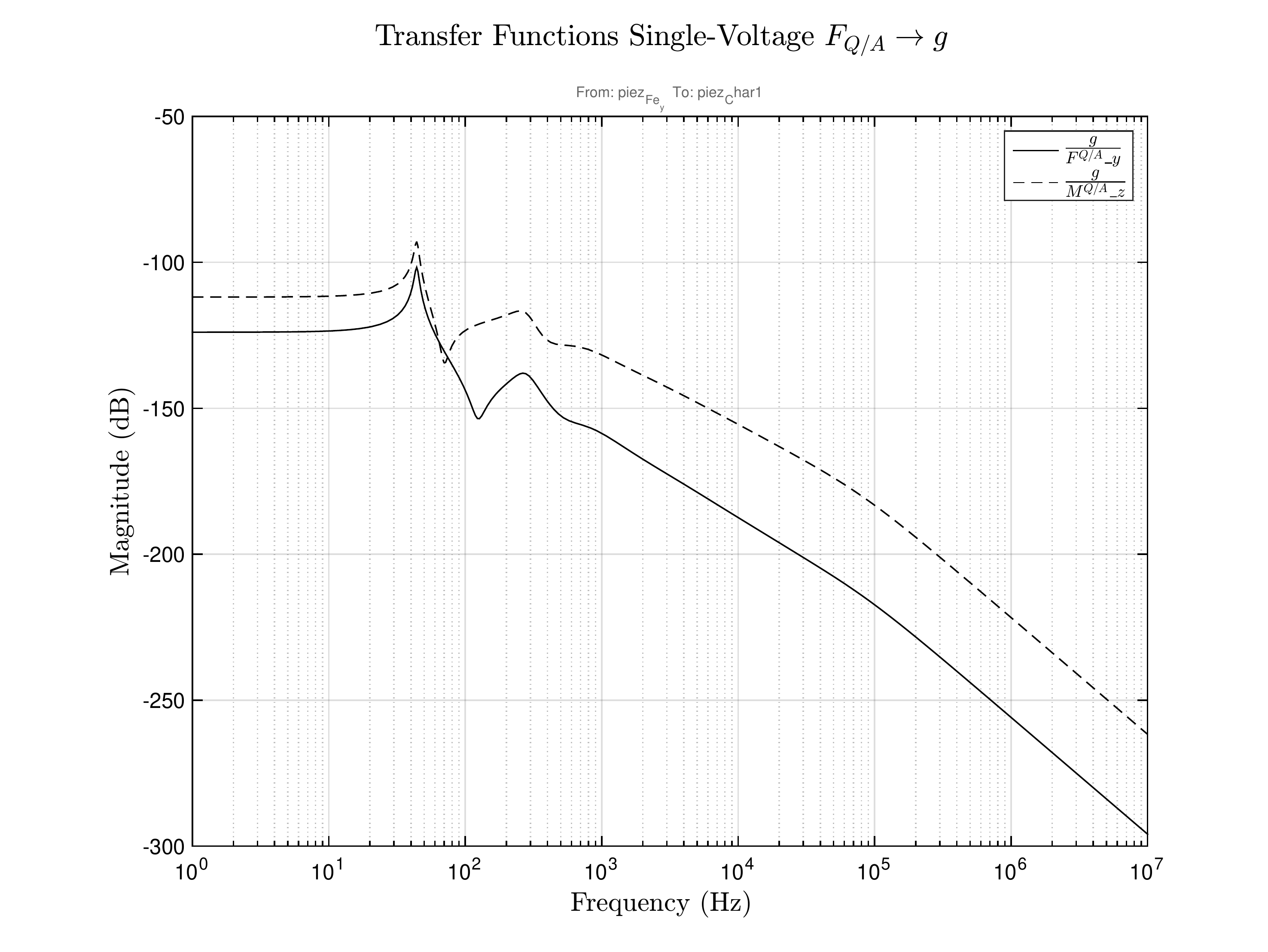}
\caption{Transfer functions between applied loads at $Q$ (vertical force and torque) and measured electric charge}
\label{fig:gVSv}
\end{figure}

\begin{figure} 
\center
\includegraphics[scale=0.3]{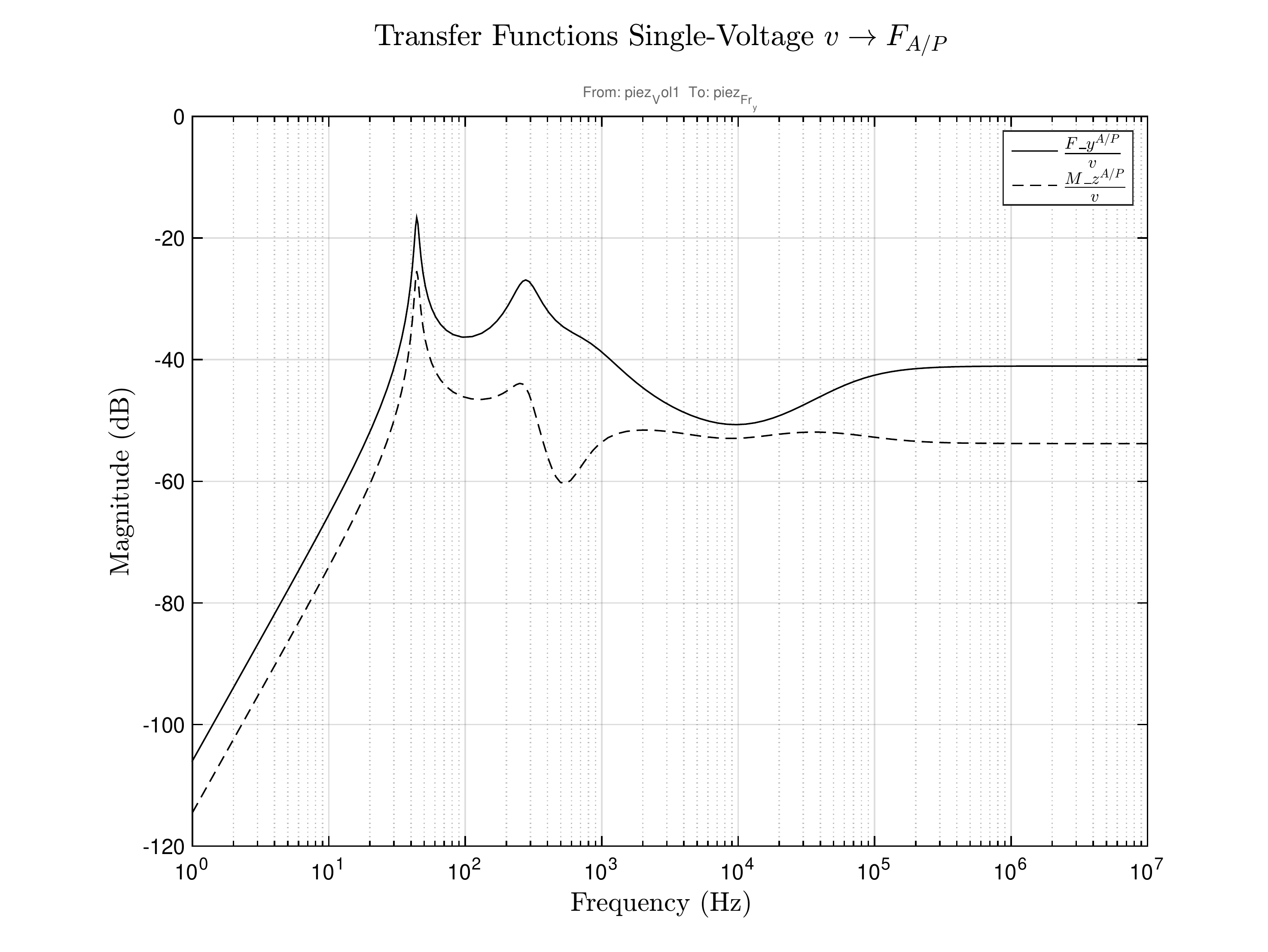}
\caption{Transfer functions between applied voltage and loads transmitted at $P$ (vertical force and torque)}
\label{fig:gVSm}
\end{figure}

After FE assembly, the complete FE model of the beam with bonded piezoelectric material is obtained. CMS is applied to the FE model as seen in Section \ref{subsec:CMS} and the TITOP model is obtained following the indications in Section \ref{subsec:SS}.  Figures \ref{fig:gVSv} and \ref{fig:gVSm} show the frequency response of a beam with piezoelectric strip discretised in 3 elements and properties extracted from the data of Table \ref{tab:parameters}. The first resonant frequency appears at 68 Hz, fast enough for the majority of large flexible systems. 

\section{Control Application}
\label{sec:control}

\begin{figure} 
 \center
\includegraphics[width = 8cm, height = 6cm]{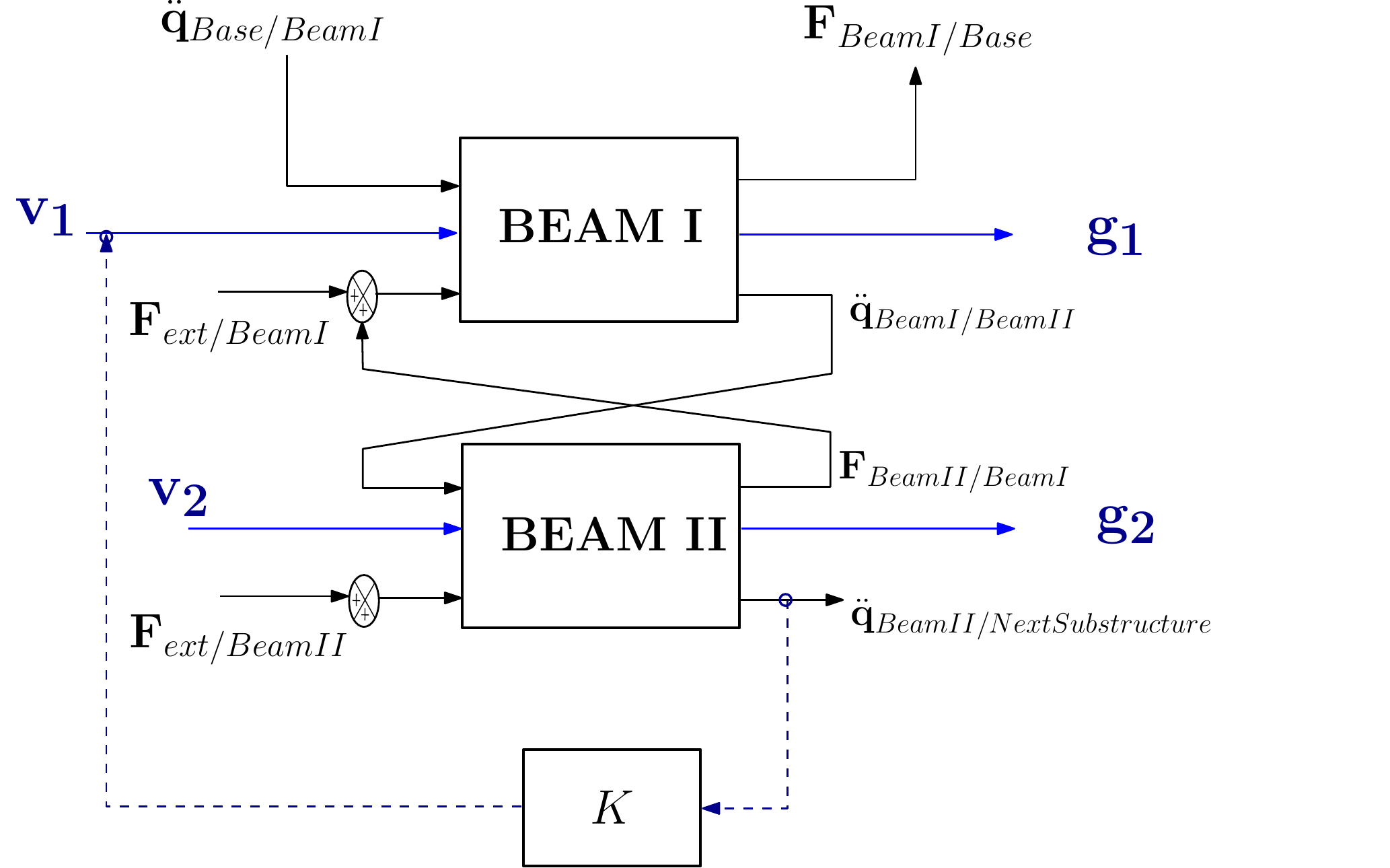}
\caption{Assembly of two flexible beams actuated through piezoelectric components with TITOP modeling technique. Rate feedback through $K$ ensures vibration attenuation.}
\label{fig:beamAssembly}
\end{figure}

As an illustrative example of applications on control of flexible structures, let consider vibration attenuation of two flexible beams in chain-like assembly with the same properties as the one modelised in Section \ref{sec:modeling}. The two flexible beams can be actuated with piezoelectric strips bonded at their surface as seen in Fig. \ref{fig:beamPiezoElement}, and their TITOP assembly is performed as depicted in Fig. \ref{fig:beamAssembly}.

\begin{figure} 
\includegraphics[width=9.3cm, height = 7cm]{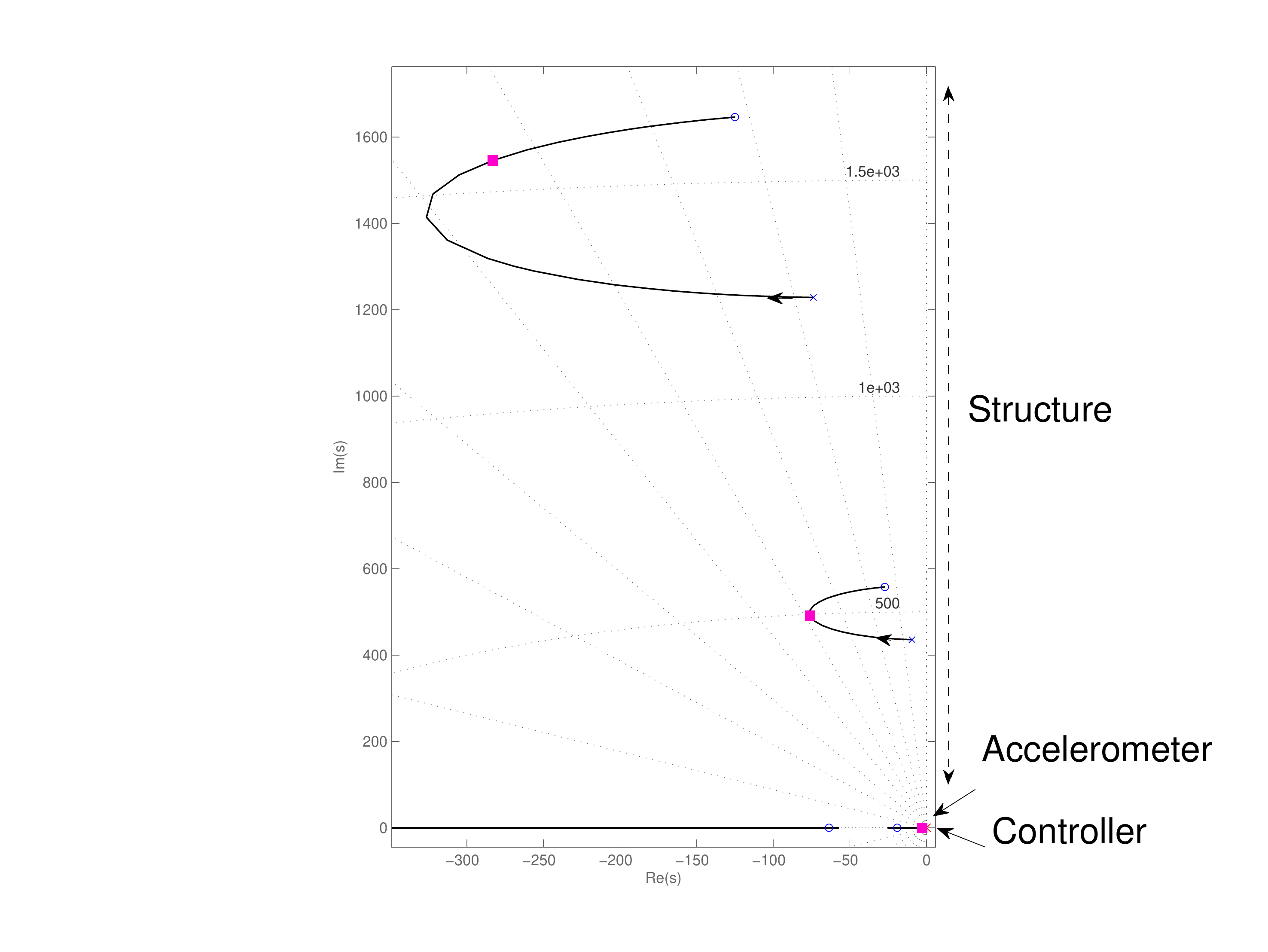}
\caption{Root locus of the rate feedback of the flexible beams. Control is targeted at mode 1}
\label{fig:rootLocus}
\end{figure}

\begin{figure} 
\center
\includegraphics[width = 8cm, height = 7cm]{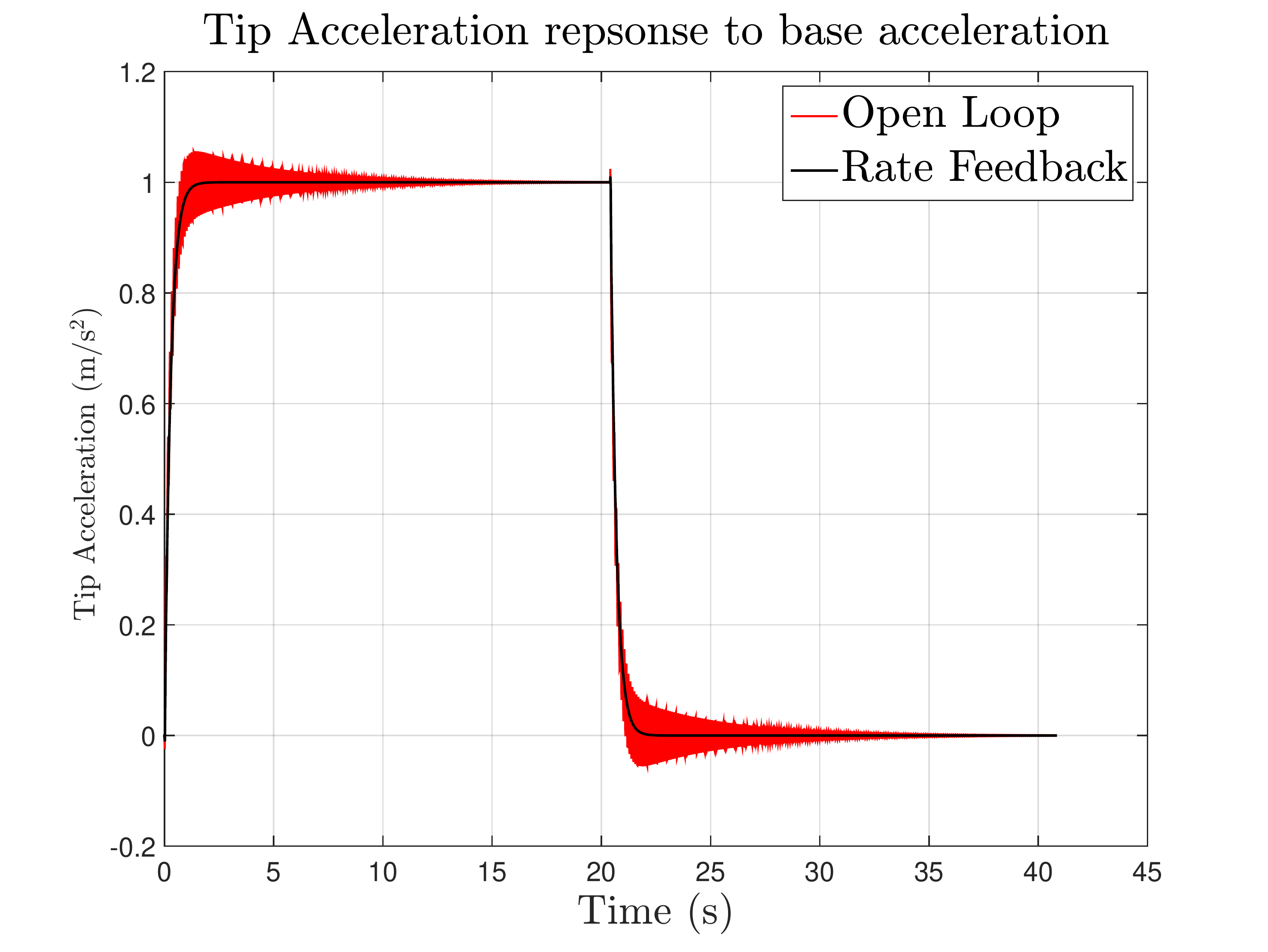}
\caption{Tip response to base acceleration in open loop and with rate feedback controller}
\label{fig:tipResponse}
\end{figure}

After assembly, vibration attenuation is addressed with a rate feeback between the integration of the acceleration at the end of Beam II, $\ddot{q}$, and the voltage applied at the piezoelectric strip in Beam I, $v_1$. Controller gain is tuned so that the first flexible mode is maximaly damped, which results in 0.2 damping ratio (see Fig. \ref{fig:rootLocus}). Simulation of the system response to base acceleration excitation shows in Fig. \ref{fig:tipResponse} that vibrations at the end of Beam II are adequately attenuated when compared with the open loop response. Other controller inputs are possible, such as electric charges $g_1$ and $g_2$ and Beam I acceleration, but Beam II rate feedback acceleration was chosen since it offered more controllability. 

Therefore, the actuated TITOP model is a straightforward modeling tool for flexible structure control applications. Indeed, this model is used for piezoelectric component modelization in a Integrated Control/Structure Design of a Large Space Structure in \cite{Perez2016_ID}.

\section{Conclusions and Perspectives}

This study has introduced a generic model for an actuated flexible structure with piezoelectric components. The model, derived from the finite element formulation, is an extension of the TITOP modeling technique, which allows interconnection with different substructures through acceleration-load transfer among the connection points. Two applications of the actuated TITOP model have been presented: modeling of a beam with bonded piezoelectric strip and vibration attenuation of a chain of actuated flexible beams. The model has resulted useful and straightforward in both applications, modeling and control.

Further research must be done in the future to improve the model including other piezoelectric effects such as histeris and creep. Other modules can be added in order to directly detect substructure displacements through electric charge measurement. This model will be used in different Integrated Control/Structure Design of Large Flexible Structures, such as in the study of \cite{Perez2016_ID}.

\begin{ack}
Research efforts were supported by the kind financial support of the CNES AOCS Department and ONERA Flight Dynamics Department in Toulouse.
\end{ack}

\bibliography{mabiblioGENERAL}             
                                                   







\appendix

\section{mass and stiffness matrices}    
\label{sec:appendix1}
The mass and stiffness matrices of each piezo element are written as:
\begin{equation}
\label{eq:m}
\begin{bmatrix}
\mathcal{M}_{qq}
\end{bmatrix}_{piezo} = \rho_pA_pl\begin{bmatrix}
 \frac{1}{3} & 0 & 0 & \frac{1}{6} & 0 & 0\\ 0 & \frac{156}{420} & \frac{22}{420}l &0  &\frac{54}{420} & -\frac{13}{420}l \\ 0 & \frac{22}{420}l & \frac{4}{420}l^2 & 0 & \frac{13}{420}l & - \frac{3}{420}l^2 \\ \frac{1}{6} & 0 & 0 & \frac{1}{3} &0 & 0 \\  0 & \frac{54}{420}l & \frac{13}{420}l & 0 & \frac{156}{420} & - \frac{22}{420}l \\ 0 & -\frac{13}{420}l & -\frac{3}{420}l^2 & 0 & -\frac{22}{420}l & \frac{4}{420}l^2
\end{bmatrix}
\end{equation}

\begin{equation}
\label{eq:k}
\begin{bmatrix}
\mathcal{K}_{qq}
\end{bmatrix}_{piezo} = \frac{E_pI_p}{l^3}\begin{bmatrix}
 \frac{A_pl^2}{I_p} & 0 & 0 & -\frac{A_pl^2}{I_p} & 0 & 0\\ 0 & 12 & 6l &0  &-12 & 6l \\ 0 & 6l & 4l^2 & 0 & -6l & 2l^2 \\ -\frac{A_pl^2}{I_p} & 0 & 0 & \frac{A_pl^2}{I_p} &0 & 0 \\  0 & -12 & -6l & 0 & 12 & -6l \\ 0 & 6l & 2l^2 & 0 & -6l & 4l^2
\end{bmatrix}
\end{equation}

where $I_p = w_pt_p(t^2 + t_pt + \frac{t_p}{2})$. The mass and stiffness matrices of each beam element are written as:
\begin{equation}
\label{eq:m}
\begin{bmatrix}
\mathcal{M}_{qq}
\end{bmatrix}_{beam} = \rho A l\begin{bmatrix}
 \frac{1}{3} & 0 & 0 & \frac{1}{6} & 0 & 0\\ 0 & \frac{156}{420} & \frac{22}{420}l &0  &\frac{54}{420} & -\frac{13}{420}l \\ 0 & \frac{22}{420}l & \frac{4}{420}l^2 & 0 & \frac{13}{420}l & - \frac{3}{420}l^2 \\ \frac{1}{6} & 0 & 0 & \frac{1}{3} &0 & 0 \\  0 & \frac{54}{420}l & \frac{13}{420}l & 0 & \frac{156}{420} & - \frac{22}{420}l \\ 0 & -\frac{13}{420}l & -\frac{3}{420}l^2 & 0 & -\frac{22}{420}l & \frac{4}{420}l^2
\end{bmatrix}
\end{equation}

\begin{equation}
\label{eq:k}
\begin{bmatrix}
\mathcal{K}_{qq}
\end{bmatrix}_{beam} = \frac{EI}{l^3}\begin{bmatrix}
 \frac{A l^2}{I} & 0 & 0 & -\frac{A l^2}{I} & 0 & 0\\ 0 & 12 & 6l &0  &-12 & 6l \\ 0 & 6l & 4l^2 & 0 & -6l & 2l^2 \\ -\frac{A l^2}{I} & 0 & 0 & \frac{A l^2}{I} &0 & 0 \\  0 & -12 & -6l & 0 & 12 & -6l \\ 0 & 6l & 2l^2 & 0 & -6l & 4l^2
\end{bmatrix}
\end{equation}

Note that the mass and stiffness matrices of the beam element are identical as the ones denoted in Eqs. (\ref{eq:m}) and (\ref{eq:k}) but substituting the piezoelectric strip parameters by those of the beam.

\section{Beam Parameters}              
                  
\begin{table}[h!]
\caption{}
\begin{center}
\label{tab:parameters}
\begin{tabular}{c c c}
& & \\ 
\hline
\hline
Beam Parameters & Symbol & Value\\
\hline
Number of Elements & $n$ & 3 \\
Total length & $L$ & 0.50 m\\
Element length & $l=L/n$ & 0.17 m\\
Thickness & $t$ & 9.53 mm \\
Width & $w$ & 30.00 mm \\
Volumetric Density & $\rho$ & 2600 Kg/m$^3$ \\
Elastic modulus & $E$ & 60 GPa \\
\hline
\hline
Piezo Parameters & Symbol & Value\\
\hline
Number of Elements & $n$ & 0.5 \\
Element length & $l=L/n$ & 0.17 m\\
Thickness & $t_p$ & 2 mm \\
Width & $w_p$ & 30 mm \\
Volumetric Density & $\rho_p$ & 7600 Kg/m$^3$ \\
Elastic modulus & $E_p$ & 50 GPa \\
Piezoelectric Constant & $d_{31}$ & -150$\times 10^{-12}$ m/V \\
Dielectric Constant & $\epsilon_{33}^T$ & 1.59$\times 10^{-12}$ F/m \\
\hline
\hline
\end{tabular}
\end{center}
\end{table}       
                                                
\end{document}